\documentclass[preprint,showpacs,preprintnumbers,amsmath,amssymb]{revtex4}

\usepackage{graphicx}
\usepackage{dcolumn}
\usepackage{bm}
\usepackage{times,mathptmx}
\usepackage{color}
\usepackage{hyperref}

\newcommand{\ave}[1]{\langle #1 \rangle}%

\begin{document}


\title{\bf Coherent mixing of mechanical excitations in nano-optomechanical structures}

\author{Qiang Lin, Jessie Rosenberg, Darrick Chang, Ryan Camacho, Matt Eichenfield, Kerry J. Vahala}
\author{Oskar Painter}
 \email{opainter@caltech.edu}
 \homepage{http://copilot.caltech.edu}
\affiliation{%
Thomas J. Watson, Sr., Laboratory of Applied Physics, California Institute of Technology, Pasadena, CA 91125}%

\date{\today}


\begin{abstract}
The combination of large per-photon optical force and small motional mass attainable in nanocavity optomechanical systems results in strong dynamical back-action between mechanical motion and the cavity light field.  In this work we study the optical control of mechanical motion within two different nanocavity structures, a zipper nanobeam photonic crystal cavity and a double-microdisk whispering-gallery resonator.  The strong optical gradient force within these cavities is shown to introduce signifcant optical rigidity into the structure, with the dressed mechanical states renormalized into optically-bright and optically-dark modes of motion.  With the addition of internal mechanical coupling between mechanical modes, a form of optically-controlled mechanical transparency is demonstrated in analogy to electromagnetically induced transparency of three-level atomic media.  Based upon these measurements, a proposal for coherently transferring RF/microwave signals between the optical field and a long-lived dark mechanical state is described.   
\end{abstract}

\maketitle

The coherent mixing of multiple excitation pathways provides the underlying mechanism for many physical phenomena. Well-known examples include the Fano resonance \cite{Fano61} and electromagnetically induced transparency (EIT) \cite{Harris90}, arising from the interference between excitations of discrete states and/or a continuum background. In the past few decades, Fano-like or EIT-like resonances have been discovered in a variety of physical systems, such as electron transport in quantum wells/dots \cite{Faist97,Kroner08}, phonon interactions in solids \cite{Scott74,Hase06}, inversion-free lasers \cite{Harris89,Imamoglu99}, coupled photonic microcavities \cite{Fan02,Boyd04,Xu06,Totsuka07}, and plasmonic metamaterials \cite{Giessen09}. Here we report a new class of coherent excitation mixing which appears in the mechanical degree of freedom of nano-optomechanical systems (NOMS). We use two canonical systems, coupled microdisks and coupled photonic-crystal nanobeams, to show that the large optical stiffening introduced by the optical gradient force actuates significant coherent mixing of mechanical excitations, not only leading to renormalization of the mechanical modes, but also producing Fano-like and EIT-like optomechanical interference, both of which are fully tunable by optical means. The demonstrated phenomena introduce the possibility for classical/quantum information processing via optomechanical systems, providing an on-chip platform for tunable optical buffering, storage, and photonic-phononic quantum state transfer.

Light forces within micro-mechanical systems have attracted considerable interest of late due to the demonstration of all-optical amplification and self-cooling of mesoscopic mechanical resonators\cite{Kippenberg05,Gigan06,Arcizet06,Kleckner06,Schliesser06}.  This technique for sensing and control of mechanical motion relies on the radiation pressure forces that build up in a mechanically compliant, high-Finesse optical cavity, resulting in strong dynamical back-action between the cavity field and mechanical motion.  More recently\cite{Povinelli051,Eichenfield07,Li08,Painter09,Lin09,Rosenberg09}, it has been realized that guided wave nanostructures can also be used to generate extremely large per-photon optical forces via the gradient optical force\cite{ref:Ashkin1}.  The combination of tailorable mechanical geometry, small motional mass, and large per-photon force in such nanostructures results in a regime of operation in which the dynamic response of the coupled optomechanical system can significantly differ from that of the bare mechanical structure.  In particular, the mechanical motion can be renormalized by the optical spring effect\cite{Braginsky77,Braginsky92,Sheard04,Zadeh07,Corbitt07,Painter09}, creating a highly anistropic, intensity-dependent effective elastic modulus of the optomechanical structure.          

\begin{figure}[btp]
\includegraphics[width=0.65\columnwidth]{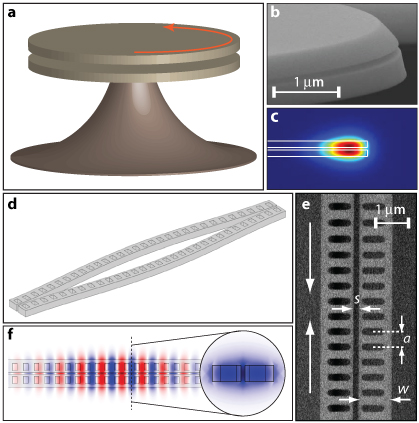}
\caption{\label{Fig1} \textbf{a}, Schematic and \textbf{b}, zoomed-in scanning electron microscopic (SEM) image of the double-disk NOMS. \textbf{c}, FEM-simulated electric field intensity of a transverse-electric (TE) polarized, bonded (even parity) whispering-gallery supermode between the two microdisks (shown in cross-section and for resonance wavelength $\lambda_c \approx 1550$~nm).  The double-disk bonded supermode has an optomechanical coupling coefficient of $g_{\text{OM}}/2\pi \approx 33$~GHz/nm.  The device studied here has a measured resonance wavelength of $\lambda_c=1538$~nm and an intrinsic and loaded quality (Q) factor of $1.07 \times 10^6$ and $0.7 \times 10^6$, respectively. \textbf{d}, Schematic, \textbf{e}, SEM image, and \textbf{f}, FEM-simulated bonded (even parity) optical supermode of the zipper cavity. The zipper cavity bonded supermode has an optomechanical coupling coefficient of $g_{\text{OM}}/2\pi \approx 68$~GHz/nm, a measured resonance wavelength of $\lambda_c=1545$~nm, and an intrinsic and loaded $Q$-factor of $3.0 \times 10^4$ and $2.8 \times 10^4$, respectively. Additional details for both devices are in the Appendices and in Refs.~\cite{Lin09,Painter09}.}
\end{figure}

In this work we focus on two specific implementations of nanoscale cavity optomechanical systems, shown in Fig.~\ref{Fig1}, in which dynamical back-action effects are particularly strong.  The first system consists of two patterned nanobeams in the near-field of each other, forming what has been termed a zipper cavity\cite{Painter09,ChanJ09}.  In this cavity structure the patterning of the nanobeams localizes light through Bragg-scattering, resulting in a series of high Finesse ($\mathcal{F}\approx 3\times 10^4$), near-infrared ($\lambda \approx 1550$~nm) optical supermodes of the beam pair.  Clamping to the substrate at either end of the suspended beams results in a fundamental in-plane mechanical beam resonance of frequency $\sim 8$ MHz.  The second cavity optomechanical system is based upon a whispering-gallery microdisk optical cavity structure.  By creating a pair of microdisks, one on top of the other with a nanoscale gap in between, strong optical gradient forces may be generated between the microdisks while maintaining the benefits of the low-loss, high-$Q$ ($Q \ge 10^6$) character of the whispering-gallery cavity.  As shown schematically in Fig.~\ref{Fig1}a, the double-disk structure\cite{Lin09} is supported and pinned at its center, allowing the perimeter of the disks to vibrate in myriad of different ways.  Of particular interest in both the zipper and double-disk systems is the differential motion of the nanobeams or disks, in which the modulation of the gap between the elements creates a large dispersive shift in the internally propagating cavity light field.  It is this type of motion that is strongly coupled to the light field, and for which the dynamical back-action is strongest.

\begin{figure}[btp]
\includegraphics[width=0.9\columnwidth]{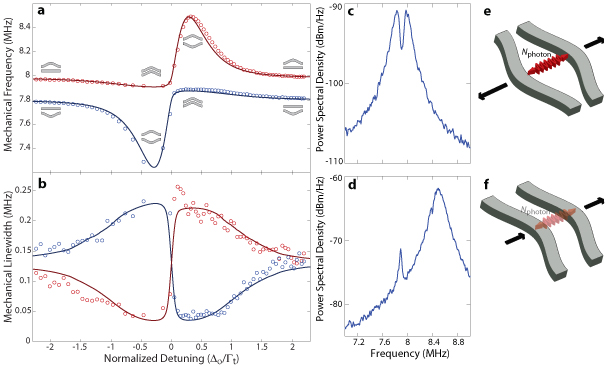}
\caption{\label{Fig2}\textbf{a}, Mechanical frequency and \textbf{b}, linewidth of the fundamental in-plane mechanical resonances of the zipper cavity's coupled nanobeams as a function of laser frequency detuning.  The input power for these measurements is 127~${\rm \mu}$W, corresponding to a maximum cavity photon number of $\sim 7000$ on resonance. The circles show the experimental data and the solid curves correspond to a fit to the data using Eq. (\ref{S_x_b_Renorm_approx}). Optically-transduced RF spectrum at a laser-cavity detuning of \textbf{c}, $\Delta_0/\Gamma_{t}=2.1$ and \textbf{d}, $\Delta_0/\Gamma_{t}=0.32$. The two nanobeams vibrate independently when the laser-cavity detuning is large, but are renormalized to the cooperative \textbf{e}, differential and \textbf{f}, common motions near resonance.}
\end{figure}

We begin with an analysis of the zipper cavity, in which the strong optically-induced rigidity associated with differential in-plane motion of the nanobeams results in a dressing of the mechanical motion by the light field.  As described in the App.~\ref{AppB} and in Ref.~\cite{Painter09}, an optical fiber nanoprobe is couple light into and out of the zipper cavity.  Optical excitation provides both a means to transduce mechanical motion (which is imparted on the transmitted light field through phase and intensity modulation) and to apply an optical-intensity-dependent mechanical rigidity via the strong optical gradient force.  By fitting a Lorentzian to the two lowest-order in-plane mechanical resonances in the radio-frequency (RF) optical transmission spectrum, we display in Fig.~\ref{Fig2}a and b the resonance frequency and resonance linewidth, respectively, of the two coupled mechanical modes of the nanobeam pair as a function of laser-cavity detuning.  At large detuning (low intra-cavity photon number) the nanobeams' motion is transduced without inducing significant optical rigidity, and the measured mechanical resonances are split by $\sim 200$ kHz, with similar linewidths (damping) and transduced amplitudes (Fig.~\ref{Fig2}(c)).  As the laser is tuned into resonance from the blue-side of the cavity, and the intra-cavity photon number increases (to $\sim 7000$), the higher frequency resonance is seen to significantly increase in frequency while the lower frequency mode tunes to the average of the independent beam frequencies with its transduced amplitude significantly weaker.  The linewidth of the high frequency resonance also tends to increase, while that of the lower frequency mode drops.  Tuning from the red-side of the cavity resonance reverses the sign of the frequency shifts and the roles of the high and low frequency modes.  

A qualitative understanding of the light-induced tuning and damping of the zipper cavity nanobeam motion emerges if one considers the effects of squeeze-film damping\cite{Bao07}.  Squeeze-film effects, a result of trapped gas in-between the beams (measurements were performed in 1 atm. of nitrogen), tend to strongly dampen differential motion of the beams and should be negligible for common motion of the beams.  Similarly, the optical gradient force acts most strongly on the differential beam motion and negligibly on the common-mode motion.  The sign of the resulting optical spring is positive for blue detuning and negative for red detuning from the cavity resonance.  Putting all of this together, a consistent picture emerges from the data in  Fig.~\ref{Fig2} in which the nanobeams start out at large detuning moving independently with similar damping (the frequency splitting of $\sim 200$ KHz is attributable to fabrication assymetries in the beams).  As the detuning is reduced, and approaches the cavity half-linewidth, the motion of the nanobeams is dressed by the internal cavity field into differential motion with a large additional optical spring constant (either positive or negative) and large squeeze-film damping component, and common motion with reduced squeeze-film damping and minimal coupling to the light field.  Due to the strong light-field coupling of the differential mode and the correspondingly weak coupling of the common mode, we term these dressed motional states \emph{optically-bright} and \emph{optically-dark}, respectively.  

A quantitative model of the dressed system can be obtained by considering the following set of coupled equations for the mechanical motion of the nanobeams:
\begin{eqnarray}
\frac{d^2 x_1}{dt^2} + \Gamma_{m} \frac{d x_1}{dt} + \Omega_{m1}^2 x_1 &=& \frac{F_{th} + F_o + F_q}{m}, \label{dx1_dt3}\\
\frac{d^2 x_2}{dt^2} + \Gamma_{m} \frac{d x_2}{dt} + \Omega_{m2}^2 x_2 &=& \frac{F_{th} - F_o - F_q}{m}, \label{dx2_dt3}
\end{eqnarray}
Here $F_o$ is the optical gradient force, $F_q$ is the viscous force from the squeeze-film effect, and $F_{th}$ is the Langevin force due to coupling of the beams to a thermal bath (further details of the properties of these three forces are described in the Appendices).  The parameters $x_j$ and $\Omega_{mj}$ ($j=1,2$) are the mechanical displacement and the resonance frequency of the uncoupled, individual beams.  We have assumed, for simplicity, that the bare mass ($m$), intrinsic mechanical damping ($\Gamma_m$), and fluctuating thermal force ($F_{th}$) are the same for each of the nanobeams.  The resulting spectral intensity of the thermally-excited optically-bright ($x_b \equiv x_1 - x_2$) differential beam motion can be shown to be (App.~\ref{AppI}),
\begin{equation}
S_{x_b}(\Omega) = \frac{k_B T \Gamma_m}{m_b} \frac{\left| L_1(\Omega) \right|^2 + \left| L_2(\Omega) \right|^2 }{\left| L_1(\Omega) L_2(\Omega) - \frac{1}{2} \left[ f_o(\Omega)/m_b + i \Gamma_q \Omega \right] \left[ L_1(\Omega) + L_2(\Omega) \right] \right|^2}, \label{S_x_b_Renorm_approx}
\end{equation}
where $L_j(\Omega) = \Omega_{mj}^2 - \Omega^2 - i \Gamma_{m} \Omega$ ($j=1,2$), $m_b = m/2$ ($\approx 10.75$ pg) is the effective mass of the differential mode, $T$ is the bath temperature, and $\Gamma_q$ represents the damping rate introduced by the squeezed gas film in between the beams.  $f_o(\Omega)$ represents the modification to the mechanical susceptibility of the differential beam motion by the optical gradient force, and is approximately given in the bad-cavity limit ($\Gamma_t \gg \Omega_m$) by,
\begin{equation}
f_{o}(\Omega) \approx - \left(2 g_{\text{OM}}^2 |a_0|^2\right) \left(\frac{\Delta_0(1+i\Gamma_{t}\Omega)}{(\Delta_0+\Omega)^2+ (\Gamma_t/2)^2}\right), \label{fo_term}
\end{equation}
where $g_{\text{OM}}$ ($\equiv d\omega_c/dx_b \approx 2\pi(68$~GHz/nm$)$) is the optomechanical coupling coefficient, $|a_0|^2$ is the time-averaged optical cavity energy, and $\Gamma_t$ ($\approx 2\pi(6.9$~GHz$)$) is the energy decay rate of the loaded optical cavity resonance.  Lorentzian fits to the double resonances of $S_{x_b}(\Omega)$ yields the frequency and damping curves shown as solid lines in Fig.~\ref{Fig2}b, displaying excellent agreement with the experimental measurements.

\begin{figure}[btp]
\includegraphics[width=0.70\columnwidth]{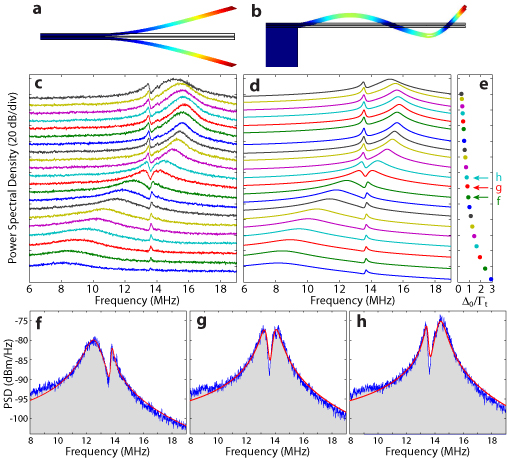}
\caption{\label{Fig3} \textbf{a,b}, FEM simulated mechanical motion of the differential flapping mode (\textbf{a}) and the common breathing mode (\textbf{b}), with simulated frequencies of 7.95 and 14.2~MHz. The color map indicates the relative magnitude (exaggerated) of the mechanical displacement. \textbf{c}, Recorded power spectral density (PSD) of the cavity transmission for the double-disk, with an input power of $315$~$\mu$W. Each curve corresponds to a normalized laser-cavity frequency detuning, $\Delta_0/\Gamma_{t}$ indicated in \textbf{e}. For display purposes, each curve is relatively shifted by 10 dB in the vertical axis. \textbf{d}, The corresponding theoretical PSD. \textbf{f-h}, Detailed PSD at three frequency detunings indicated by the arrows in \textbf{e}, with the experimental and theoretical spectra in blue and red, respectively.}
\end{figure}

A similar optically-induced renormalization mechanism applies to the double-disk cavity structure shown in Fig.~\ref{Fig1}a-c. In this case, the large optical spring effect for the differential motion of the two microdisks excites another, more intriguing form of coherent optomechanical mixing with the optically dark common mode of the disks.  Unlike in the zipper cavity, FEM modeling of the mechanics of the double-disk structure indicates a significant frequency splitting between the differential and common modes of motion of the double disk (shown in Fig.~\ref{Fig3}b and c), primarily due to the difference in the extent of the undercut between the disk layers and the extent of the central pedestal which pins the two disk layers.  The result is that the differential, or ``flapping'' motion,  of the undercut disk region has a lower frequency of $7.95$ MHz, whereas the common motion of the disks results in a higher frequency ($14.2$ MHz) ``breathing'' motion of the entire double-disk structure.   

The RF-spectrum of the transmitted optical intensity through a double-disk cavity, measured using the same fiber probing technique as for the zipper cavity (App.~\ref{AppB}), is shown in Fig.~\ref{Fig2}c versus laser-cavity detuning.  For the largest detuning (in which the optical spring is negligible) the spectrum shows a broad ($2.1$~MHz) resonance at $8.3$~MHz and a much narrower ($0.11$ MHz) resonance at $13.6$~MHz, in good corresponce with the expected frequencies of the flapping and breathing modes, respectively.  The difference in damping between the two resonances can be attributed to the strong squeeze-film damping of the differential flapping motion of the disks.  As shown in Fig.~\ref{Fig2}c, the flapping mode can be tuned in frequency via the optical spring effect from its bare value of $8.3$ MHz all the way out to $15.7$ MHz (optical input power of $P_{i}=315$~$\mu$W).  In the process, the flapping mode is tuned across the breathing mode at $13.6$ MHz. Although the optically-dark breathing mode is barely visible in the tranduced spectrum at large laser-cavity detunings, its spectral amplitude is considerably enhanced as the optically-bright flapping mode is tuned into resonance.  In addition, a strong Fano-like lineshape, with $\sim$13~dB anti-resonance, appears in the power spectrum near resonance of the two modes (Fig.~\ref{Fig3}f, g, and h).

\begin{figure}[btp]
\includegraphics[width=0.75\columnwidth]{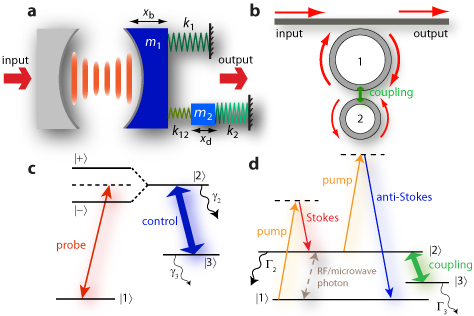}
\caption{\label{Fig4} \textbf{a}, Schematic of an equivalent Fabry-Perot cavity system showing mechanical mode mixing. The mechanical motion of the cavity mirror ($m_{1}$, equivalent to the optically-bright flapping mode) is primarily actuated by the spring $k_1$ and the optical force. It is internally coupled to a second mass-spring system ($m_{2}$, equivalent to the breathing mode) actuated by the spring $k_2$ which is decoupled from the optical wave. The two masses are internally coupled via spring $k_{12}$.  \textbf{b}, A photonic analogue to the optomechanical system involving coupled resonators. Microcavity 1 is directly coupled to the external optical waveguide (equivalent to the optically-bright flapping mode) and also internally coupled the narrowband cavity 2 (equivalent to the optically-dark breathing mode).  \textbf{c}, State diagram of an EIT-like medium. The excited state ($|2\rangle$) is split by the optical control beam into two broadband dressed states ($|+\rangle$ and $|-\rangle$). The dipole transition between ground-states $|1\rangle$ and $|3\rangle$ is forbidden. \textbf{d}, The state diagram corresponding to the optomechanical system of \textbf{a}, where $|1\rangle$ is the phonon vacuum state, and $| 2 \rangle$ and $| 3 \rangle$ correspond to the flapping and breathing modes, respectively.}
\end{figure}

As shown schematically in Fig.~\ref{Fig4}(a), the Fano-like interference in the optically-bright power spectral density can be attributed to an internal mechanical coupling between the flapping and breathing mechanical modes. This is quite similar to the phonon-phonon interaction during the structural phase transition in solids \cite{Scott74, Barker64, Porto68, Scott70, Zawadowski70, Porto74}, in which the internal coupling between phonon modes produces Fano-like resonances in the Raman-scattering spectra.  A Hamiltonian for the coupled optomechanical system can be written as,
\begin{equation}
\mathcal{H}_m = \frac{p_b^2}{2 m_b} + \frac{1}{2} m_b \Omega_{mb}^2 x_b^2 + \frac{p_d^2}{2 m_d} + \frac{1}{2} m_d \Omega_{md}^2 x_d^2 + \kappa x_b x_d, \label{H_m}
\end{equation}
where $x_j$, $p_j$, $\Omega_j$, and $m_j$ ($j=b,d$) are the mechanical displacement, kinetic momentum, intrinsic mechanical frequency, and effective motional mass, respectively, for the optically-bright flapping ($j=b$) and optically-dark breathing ($j=d$) mechanical modes. $\kappa$ represents the internal mechanical coupling between the two modes. Equation (\ref{H_m}) together with the three external actuation forces (optical, squeeze-film, and thermal in nature) leads to the following equations of motion for the two mechanical modes:
\begin{eqnarray}
\frac{d^2 x_b}{dt^2} + \Gamma_{mb} \frac{d x_b}{dt} + \Omega_{mb}^2 x_b + \frac{\kappa}{m_b} x_d &=& \frac{F_b}{m_b} + \frac{F_o}{m_b}, \label{dx1_dt}\\
\frac{d^2 x_d}{dt^2} + \Gamma_{md} \frac{d x_d}{dt} + \Omega_{md}^2 x_d + \frac{\kappa}{m_d} x_b &=& \frac{F_d}{m_d}, \label{dx2_dt}
\end{eqnarray}
where $F_j$ ($j=b,d$) represents the thermal Langevin forces and we have incorporated the squeeze-film damping into $\Gamma_{mb}$ for the optically-bright differential motion. Equations (\ref{dx1_dt}) and (\ref{dx2_dt}) lead to a spectral intensity for $x_b$ of (App.~\ref{AppE}):
\begin{equation}
S_{x_b}(\Omega) = \frac{2 k_B T}{m_b} \frac{\eta^4 \Gamma_{md} + \Gamma_{mb} \left| L_d(\Omega) \right|^2}{\left| L_b(\Omega) L_d(\Omega) - \eta^4 \right|^2}, \label{S_x1}
\end{equation}
where $\eta^4 \equiv \kappa^2/(m_b m_d)$ represents the mechanical coupling coefficient, $L_b(\Omega) = \Omega^2 - \Omega_{mb}^2 - i \Gamma_{mb} \Omega - f_o(\Omega)/m_b$, and $L_d(\Omega) = \Omega^2 - \Omega_{md}^2 - i \Gamma_{md} \Omega$.  As discussed in more detail in Ref. \cite{Lin09}, the parameters for the double-disk are $m_b=264$~pg, $g_{\text{OM}}/2\pi=33$~GHz/nm, and $\Gamma_{t}/2\pi=280$~MHz.  The power spectral density of the cavity transmission, which we measure optically, is linearly proportional to $S_{x_b}(\Omega)$. As shown in Fig.~\ref{Fig2}d, the theoretical model (with fitting parameter $\eta/2\pi = 3.32$~MHz) provides an excellent description of the observed phenomena.  In App.~\ref{AppH} we present similar coherent mode-mixing involving two optically-dark modes and a single optically-bright mode in the zipper cavity. 

The mechanical response given by Eq.~(\ref{S_x1}) is directly analogous to the atomic response in EIT~\cite{Imamoglu05}.  Just as in EIT, one can understand the resulting Fano lineshape in two different ways.  The first perspective considers the interference associated with multiple excitation pathways.  In the optomechanical system, the mechanical motion of the flapping mode is thermally excited along two different pathways, either directly into the broadband (lossy) flapping mode, or indirectly, through the flapping mode, into the long-lived breathing mode, and then back again into the flapping mode. The two excitation pathways interfere with each other, resulting in the Fano-like resonance in the spectral response of the optically bright flapping mode.  An alternative, but perfectly equivalent view of the coupled optomechanical system considers the dressed states resulting from the internal mechanical coupling.  In this picture the internal mechanical coupling renormalizes the broadband flapping mode and the narrowband breathing mode into two dressed mechanical modes, both broadband and optically-bright. In particular, when the flapping and breathing mechanical frequencies coincide, the two dressed modes are excited with equal amplitude and opposite phase at the center frequency between the split dressed states. Destructive interference results, suppressing excitation of the mechanical system at the line center. Consequently, the mechanical motion becomes purely a trapped \emph{mechanically-dark} state, transparent to external excitation. As shown in Fig.~\ref{Fig3}c, this induced mechanical transparency is a direct analogue to EIT in atomic systems \cite{Harris90,Imamoglu05,Alzar02, Hemmer88}, in which the quantum interference between the transition pathways to the dressed states of the excited electronic state, through either $| 1 \rangle \leftrightarrow | + \rangle$ or $| 1 \rangle \leftrightarrow | - \rangle$, leads to an induced spectral window of optical transparency.

Despite the intriguing similarities between the optomechanical system studied here and EIT in atomic media, there are some important, subtle differences.  For instance, in the optomechanical system, rather than the linear dipole transition of EIT, the interaction corresponds to a second-order transition. The dynamic backaction between the cavity field and mechanical motion creates Stokes and anti-Stokes optical sidebands, whose beating with the fundamental optical wave resonates with the mechanical motion to create/annihilate phonons (see Fig.~\ref{Fig3}d).  Functionally, this is like coherent Stokes and anti-Stokes Raman scattering, albeit with unbalanced scattering amplitudes resulting from the coloring of the electromagnetic density of states by the optical cavity (App.~\ref{AppG}). Therefore, in analogy to EIT, it is the modulation signal carried by the incident optical wave (radio-frequency or microwave photons) that fundamentally probes/excites the mechanical motion and to which the trapped \emph{mechanically-dark} state becomes transparent. Moreover, rather than tuning the Rabi-splitting through the intensity of a control beam resonant with the $|3 \rangle \leftrightarrow |2 \rangle$ electronic transition (Fig.~\ref{Fig3}c), this \emph{optically-induced mechanical transparency} is controlled via optical spring tuning of the resonance frequency of the optically bright flapping mechanical mode.  Perhaps the most apt analogy to the optomechanical system can be made to the photonic resonator system shown in Fig.~\ref{Fig3}d.  The interference in this case is between the two optical pathways composed of the waveguide-coupled low-$Q$ optical resonator 1, and the waveguide-decoupled high-$Q$ resonator 2.  This interference again leads to a Fano-like resonance, or what has been termed coupled-resonator-induced transparency, in the optical cavity transmission~\cite{Fan02, Boyd04, Xu06, Totsuka07}. 

Although the studies considered here involve thermal excitation of the optomechanical system, the same phenomena can be excited more efficiently, and with greater control, using external optical means (App.~\ref{AppF}).  As such, beyond the interesting physics of these devices, exciting application in RF/microwave photonics and quantum optomechanics exist.  Similar to the information storage realized through EIT~\cite{Imamoglu05, Hau01, Lukin03}, optical information can be stored and buffered in the dark mechanical degree of freedom in the demonstrated NOMS. This can be realized through a procedure similar to that recently proposed for coupled optical resonators~\cite{Fan04, Fan05} in which dynamic, adiabatic tuning of optical resonances are used to slow, store, and retrieve optical pulses.  The corresponding optomechanical system would consist of an array of double-disk resonators, all coupled to a common optical bus waveguide into which an optical signal carrying RF/microwave information would be launched.  In this scheme, a second control optical beam would adiabatically tune the frequency of the optically-bright flapping mode of each resonator, allowing for the RF/microwave signal to be coherently stored in (released from) the long-lived breathing mode through adiabatic compression (expansion) of the mechanical bandwidth~\cite{Fan04,Fan05}.  In comparison to the all-photonic system, optomechanical systems have several advantages, primarily related to the attainable lifetime of the dark mechanical state.  For example, the radial breathing mechanical mode of a similar whispering-gallery cavity has been shown to exhibit a lifetime of more than 2 ms~\cite{Anetsberger08}, a timescale more than seven orders of magnitude longer than that in demonstrated photonic-coupled-resonator systems~\cite{Xu07} and comparable with EIT media~\cite{Hau01,Lukin03}.  Moreover, mechanical lifetimes of more than one second have recently been demonstrated using stressed silicon nitride nanobeam~\cite{Verbridge081} and nanomembrane~\cite{ThompsonJD08} mechanical resonators operating in the MHz frequency regime.  In the quantum realm, such a system operating in the good-cavity or sideband-resolved regime (by increasing either the optical $Q$ factor \cite{Kippenberg03} or the mechanical frequency), would reduce the \emph{simultaneous} creation and annihilation of Stokes and anti-Stokes photons, enabling efficient information storage and retrieval at the single-quanta-level suitable for quantum state transfer.

\section*{Acknowledgements}The authors would like to thank Thomas Johnson and Raviv Perahia for their help with the device processing, and Thiago Alegre for helpful discussions.  This work was supported by the NSF (EMT grant no. 0622246 and CIAN grant no. EEC-0812072 through University of Arizona) and through a seedling program from DARPA (grant no. HR0011-08-0002).


\appendix

\section{Device fabrication}
\label{AppA}
The zipper cavity is formed from a thin-film ($400$~nm) of tensile-stressed, stochiometric Si$_{3}$N$_{4}$ deposited by low-pressure chemical vapor deposition on a silicon substrate.  Electron beam-lithography, followed by a series of plasma and wet chemical etches, are used to form the released nanobeam structure.  The double-disk structure is formed from a $158$~nm sacrifical amorphous silicon layer sandwiched in between two $340$~nm thick silica glass layers, all of which are deposited via plasma-enhanced chemical vapor deposition.  A high temperature ($1050$~K) thermal anneal is used to improve the optical quality of the as-deposited silica layers.  The microdisk pattern was fabricated by reactive ion etching, and the sandwiched $\alpha$-Si layer was undercut by 6~${\rm \mu}$m from the disk edge using a sulfur hexafluoride dry release etch.  This etch simultaneously undercuts the silicon substrate to form the underlying silicon pedestal. The final air-gap between the silica disks size is measured to be $138$~nm due to shrinkage of the amorphous silicon layer during annealing.   

\section{Optical probing and RF spectrum measurements}
\label{AppB}
A fiber-taper optical coupling technique is used to in-couple and out-couple light from the zipper and double-disk cavities.  The fiber taper, with extremely low-loss ($88 \%$ transmission efficiency), is put in contact with the substrate near the cavities in order to mechanically anchor it during all measurements (thus avoiding power-dependent movement of the taper due to thermal and/or optical forces).  An optical fiber polarization controller, consisting of a series of circular loops of fiber, is used to selectively excite the transverse-electric polarized optical modes of both cavities. 

RF spectra are measured by direct detection of the optical power transmitted through the cavities using a $125$ MHz bandwidth photoreceiver (noise-equivalent-power NEP$=2.5$ pW/Hz$^{1/2}$ from $0$-$10$ MHz and $22.5$ pW/Hz$^{1/2}$ from $10$-$200$ MHz, responsivity $R = 1$ A/W, transimpedance gain $G=4\times10^4$ V/A) and a high-speed oscilloscope ($2$ Gs/s sampling rate and $1$ GHz bandwidth).  A pair of ``dueling'' calibrated optical attenuators are used before and after the cavities in order to vary the input power to the cavity while keeping the detected optical power level constant.  The measured electrical noise floor is set by the circuit noise of the photodetector for the optical power levels considered in this work, corresponding to $-125$ dBm/Hz near $10$ MHz.

\section{Intracavity field in the presence of optomechanical coupling}
\label{AppC}
In the presence of optomechanical coupling, the optical field inside the cavity satisfies the following equation:
\begin{equation}
\frac{da}{dt} = (i \Delta_0 - \Gamma_t/2 - i g_{om}x_b ) a + i \sqrt{\Gamma_e} A_{in}, \label{da_dt}
\end{equation}
where $a$ is the optical field of the cavity mode, normalized such that $U = |a|^2$ represents the mode energy, and $A_{in}$ is the input optical wave, normalized such that $P_{in}=|A_{in}|^2$ represents the input power. $\Gamma_t$ is the photon decay rate for the loaded cavity and $\Gamma_e$ is the photon escape rate associated with the external coupling. $\Delta_0 = \omega - \omega_0$ is the frequency detuning from the input wave to the cavity resonance. $g_{om}$ is the optomechanical coupling coefficient associated with the optically bright mode, with a mechanical displacement given by $x_b$. In Eq.~(\ref{da_dt}), we have neglected the optomechanical coupling to the optically dark mode because of its negligible magnitude.

Well below the threshold of mechanical oscillation, the mechanical motion is generally small, and its impact on the intracavity optical field can be treated as a small perturbation. As a result, the intracavity field can be written as $a(t) \approx a_0(t) + \delta a(t)$, where $a_0$ is the cavity field in the absence of optomechanical coupling and $\delta a$ is the perturbation induced by the mechanical motion. They satisfy the following two equations:
\begin{eqnarray}
\frac{da_0}{dt} &=& (i \Delta_0 - \Gamma_t/2) a_0 + i \sqrt{\Gamma_e} A_{in}, \label{da0_dt}\\
\frac{d \delta a}{dt} &=& (i \Delta_0 - \Gamma_t/2) \delta a - i g_{om}x_b a_0. \label{dDa_dt}
\end{eqnarray}
In the case of a continuous-wave input, Eq.~(\ref{da0_dt}) leads to a steady state given by
\begin{equation}
a_0 = \frac{i \sqrt{\Gamma_e} A_{in}}{\Gamma_t/2 - i \Delta_0},
\end{equation}
and Eq.~(\ref{dDa_dt}) provides a spectral response for the perturbed field amplitude of
\begin{equation}
\delta \widetilde{a}(\Omega) = \frac{i g_{om} a_0 \widetilde{x}_b(\Omega)}{i(\Delta_0 + \Omega) - \Gamma_t/2}, \label{Da_Omega}
\end{equation}
where $\delta \widetilde{a}(\Omega)$ is the Fourier transform of $\delta a(t)$ defined as $\delta \widetilde{a}(\Omega) = \int_{-\infty}^{+\infty} {\delta a(t) e^{i\Omega t} dt}$. Similarly, $\widetilde{x}_b(\Omega)$ is the Fourier transform of $x_b(t)$.

\section{The power spectral density of the cavity transmission}
\label{AppD}
From the discussion in the previous section, the transmitted optical power from the cavity is given by
\begin{equation}
P_T = \left| A_{in} + i \sqrt{\Gamma_e} a \right|^2 \approx |A_0|^2 + i \sqrt{\Gamma_e} \left( A_0^* \delta a - A_0 \delta a^* \right), \label{P_T}
\end{equation}
where $A_0$ is the steady-state cavity transmission in the absence of optomechanical coupling. It is given by
\begin{equation}
A_0 = A_{in} \frac{(\Gamma_0 - \Gamma_e)/2 - i \Delta_0}{\Gamma_t/2 - i \Delta_0}, \label{A_0}
\end{equation}
where $\Gamma_0$ is the photon decay rate of the intrinsic cavity. It is easy to show that the averaged cavity transmission is given by $\ave{P_T} = |A_0|^2$, as expected. By using Eqs.~(\ref{Da_Omega}), (\ref{P_T}), and (\ref{A_0}), we find the power fluctuations, $\delta P_T (t) \equiv P_T(t) - \ave{P_T}$, are given in the frequency domain by
\begin{equation}
\delta \widetilde{P}_T(\Omega) = \frac{i \Gamma_e P_{in}  g_{om} \widetilde{x}_b(\Omega)}{(\Gamma_t/2)^2 + \Delta_0^2} \left[ \frac{(\Gamma_0-\Gamma_e)/2 + i \Delta_0}{\Gamma_t/2 - i(\Delta_0+\Omega)} - \frac{(\Gamma_0 - \Gamma_e)/2 - i \Delta_0}{\Gamma_t/2 + i (\Delta_0 - \Omega)} \right], \label{DP_Omega}
\end{equation}
where $\delta \widetilde{P}_T(\Omega)$ is the Fourier transform of $\delta P_T(t)$. By using Eq.~(\ref{DP_Omega}), we obtain a power spectral density (PSD) for the cavity transmission of
\begin{equation}
S_P(\Omega) = g_{om}^2 P_{in}^2 S_{x_b}(\Omega) H(\Omega), \label{S_P}
\end{equation}
where $S_{x_b}(\Omega)$ is the spectral intensity of the mechanical displacement for the optically bright mode which will be discussed in detail in the following sections. $H(\Omega)$ is the cavity transfer function defined as
\begin{equation}
H(\Omega) \equiv \frac{\Gamma_e^2}{\left[ \Delta_0^2 + (\Gamma_t/2)^2 \right]^2} \frac{4 \Delta_0^2 (\Gamma_0^2 + \Omega^2)}{\left[ (\Delta_0+\Omega)^2 +(\Gamma_t/2)^2 \right] \left[ (\Delta_0 - \Omega)^2 + (\Gamma_t/2)^2 \right]}. \label{H_Omega}
\end{equation}
In general, when compared with $S_{x_b}(\Omega)$, $H(\Omega)$ is a slowly varying function of $\Omega$ and can be well approximated by its value at the mechanical resonance: $H(\Omega) \approx H(\Omega_{mb}')$. Clearly then, the power spectral density of the cavity transmission is linearly proportional to the spectral intensity of the mechanical displacement of the optically bright mode.

\section{The mechanical response with multiple excitation pathways}
\label{AppE}
When the optically bright mode is coupled to an optically dark mode, the Hamiltonian for the coupled mechanical system is given by the general form:
\begin{equation}
{\cal H}_m = \frac{p_b^2}{2 m_b} + \frac{1}{2} k_b x_b^2 + \frac{p_d^2}{2 m_d} + \frac{1}{2} k_d x_d^2 + \kappa x_b x_d, \label{H_m}
\end{equation}
where $x_j$, $p_j$, $k_j$, and $m_j$ ($j=b,d$) are the mechanical displacement, kinetic momentum, the spring constant, and the effective motional mass for the $j^{\rm th}$ mechanical mode, respectively, and $\kappa$ represents the mechanical coupling between the bright and dark modes. The subscripts $b$ and $d$ denote the optically bright and optically dark modes, respectively. With this system Hamiltonian, including the optical gradient force on the optically bright mode and counting in the mechanical dissipation induced by the thermal mechanical reservoir, we obtain the equations of motion for the two mechanical modes:
\begin{eqnarray}
\frac{d^2 x_b}{dt^2} + \Gamma_{mb} \frac{d x_b}{dt} + \Omega_{mb}^2 x_b + \frac{\kappa}{m_b} x_d &=& \frac{F_b}{m_b} + \frac{F_o}{m_b}, \label{dx1_dt}\\
\frac{d^2 x_d}{dt^2} + \Gamma_{md} \frac{d x_d}{dt} + \Omega_{md}^2 x_d + \frac{\kappa}{m_d} x_b &=& \frac{F_d}{m_d}, \label{dx2_dt}
\end{eqnarray}
where $\Omega_{mj}^2 \equiv \frac{k_j}{m_j}$ is the mechanical frequency for the $j^{\rm th}$ mode. $F_j$ ($j=b,d$) represents the Langevin forces from the thermal reservoir actuating the Brownian motion, with the following statistical properties in the frequency domain:
\begin{equation}
\ave{ \widetilde{F}_i (\Omega_u) \widetilde{F}_j^* (\Omega_v) } = 2 m_i \Gamma_{mi} k_B T \delta_{ij} 2 \pi \delta(\Omega_u - \Omega_v), \label{F_corr}
\end{equation}
where $i, j =b,d$, $T$ is the temperature and $k_B$ is the Boltzmann constant. $\widetilde{F}_i(\Omega)$ is the Fourier transform of $F_i(t)$.

In Eq.~(\ref{dx1_dt}), $F_o = - \frac{g_{om} |a|^2}{\omega_0}$ represents the optical gradient force. From the previous section, we find that it is given by
\begin{equation}
F_o(t) = -\frac{g_{om}}{\omega_0} \left[ |a_0|^2 + a_0^* \delta a(t) + a_0 \delta a^*(t)\right]. \label{F_0}
\end{equation}
The first term is a static term which only changes the equilibrium position of the mechanical motion. It can be removed simply by shifting the mechanical displacement to be centered at the new equilibrium position. Therefore, we neglect this term in the following discussion. The second and third terms provide the dynamic optomechanical coupling. From Eq.~(\ref{Da_Omega}), the gradient force is found to be given in the frequency domain by
\begin{equation}
\widetilde{F}_o(\Omega) \equiv f_o(\Omega) \widetilde{x}_b(\Omega) = - \frac{2 g_{om}^2 |a_0|^2 \Delta_0 \widetilde{x}_b(\Omega)}{\omega_0} \frac{\Delta_0^2 - \Omega^2 + (\Gamma_t/2)^2 + i \Gamma_t \Omega}{\left[(\Delta_0+\Omega)^2+ (\Gamma_t/2)^2 \right] \left[(\Delta_0 - \Omega)^2 + (\Gamma_t/2)^2 \right]}, \label{F0_Omega}
\end{equation}
which is linearly proportional to the mechanical displacement of the optically bright mode.

Equations (\ref{dx1_dt}) and (\ref{dx2_dt}) can be solved easily in the frequency domain, in which the two equations become
\begin{eqnarray}
L_b(\Omega) \widetilde{x}_b + \frac{\kappa}{m_b} \widetilde{x}_d &=& \frac{\widetilde{F}_b}{m_b} + \frac{\widetilde{F}_o}{m_b}, \label{dx1_Omega}\\
L_d(\Omega) \widetilde{x}_d + \frac{\kappa}{m_d} \widetilde{x}_b &=& \frac{\widetilde{F}_d}{m_d}, \label{dx2_Omega}
\end{eqnarray}
where $L_j(\Omega) \equiv \Omega_{mj}^2 - \Omega^2 - i \Gamma_{mj} \Omega$ ($j=b,d$). Substituting Eq.~(\ref{F0_Omega}) into Eq.~(\ref{dx1_Omega}), we find that Eq.~(\ref{dx1_Omega}) can be written in the simple form,
\begin{equation}
L_b(\Omega) \widetilde{x}_b + \frac{\kappa}{m_b} \widetilde{x}_d = \frac{\widetilde{F}_b}{m_b}, \label{dx1_Omega2}
\end{equation}
where $L_b(\Omega)$ is now defined with a new mechanical frequency $\Omega_{mb}'$ and energy decay rate $\Gamma_{mb}'$ as
\begin{equation}
L_b(\Omega) = \Omega_{mb}^2 - \Omega^2 - i \Gamma_{mb} \Omega - \frac{f_o(\Omega)}{m_b} \equiv (\Omega_{mb}')^2 - \Omega^2 - i \Gamma_{mb}' \Omega, \label{L1_New}
\end{equation}
and the new $\Omega_{mS}'$ and $\Gamma_{mS}'$ are given by
\begin{eqnarray}
(\Omega_{mb}')^2 &\equiv& \Omega_{mb}^2 + \frac{2 g_{om}^2 |a_0|^2 \Delta_0}{m_b \omega_0} \frac{\Delta_0^2 - \Omega^2 + (\Gamma_t/2)^2}{\left[(\Delta_0+\Omega)^2+ (\Gamma_t/2)^2 \right] \left[(\Delta_0 - \Omega)^2 + (\Gamma_t/2)^2 \right]} \nonumber \\
&\approx& \Omega_{mb}^2 + \frac{2 g_{om}^2 |a_0|^2 \Delta_0}{m_b \omega_0} \frac{\Delta_0^2 - \Omega_{mb}^2 + (\Gamma_t/2)^2}{\left[(\Delta_0+\Omega_{mb})^2+ (\Gamma_t/2)^2 \right] \left[(\Delta_0 - \Omega_{mb})^2 + (\Gamma_t/2)^2 \right]}, \label{Omega_mNew}\\
\Gamma_{mb}' &\equiv& \Gamma_{mb} - \frac{2 g_{om}^2 |a_0|^2 \Gamma_t \Delta_0}{m_b \omega_0} \frac{1}{\left[(\Delta_0+\Omega)^2+ (\Gamma_t/2)^2 \right] \left[(\Delta_0 - \Omega)^2 + (\Gamma_t/2)^2 \right]} \nonumber\\
&\approx& \Gamma_{mb} - \frac{2 g_{om}^2 |a_0|^2 \Gamma_t \Delta_0}{m_b \omega_0} \frac{1}{\left[(\Delta_0+\Omega_{mb})^2+ (\Gamma_t/2)^2 \right] \left[(\Delta_0 - \Omega_{mb})^2 + (\Gamma_t/2)^2 \right]}. \label{Gamma_mNew}
\end{eqnarray}
Clearly, the effect of the optical gradient force on the optically bright mode is primarily to change its mechanical frequency (the optical spring effect) and energy decay rate (mechanical amplification or damping).

Equations (\ref{dx2_Omega}) and (\ref{dx1_Omega2}) can be solved easily to obtain the solution for the optically bright mode,
\begin{equation}
\widetilde{x}_b(\Omega) = \frac{\frac{\widetilde{F}_b(\Omega)}{m_b} L_d(\Omega) - \frac{\kappa}{m_b} \frac{\widetilde{F}_d(\Omega)}{m_d}}{L_b(\Omega) L_d(\Omega) - \eta^4}, \label{x1_Omega}
\end{equation}
where $\eta^4 \equiv \frac{\kappa^2}{m_b m_d}$ represents the mechanical coupling coefficient. By using Eq.~(\ref{F_corr}) and (\ref{x1_Omega}), we obtain the spectral intensity of the mechanical displacement for the optically bright mode,
\begin{equation}
S_{x_b}(\Omega) = \frac{2 k_B T}{m_b} \frac{\eta^4 \Gamma_{md} + \Gamma_{mb} \left| L_d(\Omega) \right|^2}{\left| L_b(\Omega) L_d(\Omega) - \eta^4 \right|^2}, \label{S_x1}
\end{equation}
where $L_b(\Omega)$ is given by Eq.~(\ref{L1_New}). The mechanical response given by Eq.~(\ref{S_x1}) is very similar to the atomic response in EIT.

As discussed in the previous section, the power spectral density (PSD) of the cavity transmission is linearly proportional to Eq.~(\ref{S_x1}). Equation (\ref{S_x1}) together with (\ref{S_P}) is used to find the theoretical PSD shown in Fig.~3, by using an optomechanical coupling coefficient of $g_{om}/2\pi = 33~{\rm GHz/nm}$ and an effective mass of $m_{b} = 264$~pg for the flapping mode, both obtained from FEM simulations. The intrinsic and loaded optical quality factors of $1.07 \times 10^6$ and $0.7 \times 10^6$ are obtained from optical characterization of the cavity resonance, and are also given in the caption of Fig.~1 of the main text. The intrinsic mechanical frequencies and damping rates of the two modes ($\Omega_{mb}$, $\Omega_{md}$, $\Gamma_{mb}$, and $\Gamma_{md}$) are obtained from the experimentally recorded PSD of cavity transmission with a large laser-cavity detuning, as given in the caption of Fig.~3 of the main text. The mechanical coupling coefficient $\eta$ is treated as a fitting parameter. Fitting of the PSDs results in $\eta = 3.32$~MHz, indicating a strong internal coupling between the two mechanical modes. As shown clearly in Fig.~3d, f-h of the main text, our theory provides an excellent description of the observed phenomena.

\section{The mechanical response with external optical excitation}
\label{AppF}
The previous section focuses on the case in which the mechanical excitations are primarily introduced by the thermal perturbations from the environmental reservoir. However, the mechanical motion can be excited more intensely through the optical force by modulating the incident optical wave. In this case, the input optical wave is composed of an intense CW beam together with a small modulation: $A_{in} = A_{in0} + \delta A(t)$. As a result, Eq.~(\ref{dDa_dt}) now becomes
\begin{equation}
\frac{d \delta a}{dt} = (i \Delta_0 - \Gamma_t/2) \delta a - i g_{om}x_b a_0 + i \sqrt{\Gamma_e} \delta A. \label{dDa_dt_Modulated}
\end{equation}
This equation leads to the intracavity field modulation given in the frequency domain as:
\begin{equation}
\delta \widetilde{a}(\Omega) = \frac{i g_{om} a_0 \widetilde{x}_b(\Omega) - i \sqrt{\Gamma_e} \delta \widetilde{A}(\Omega)}{i(\Delta_0 + \Omega) - \Gamma_t/2}, \label{Da_Omega_Modulated}
\end{equation}
where $\delta \widetilde{A}(\Omega)$ is the Fourier transform of $\delta A (t)$. By use of this solution together with Eq.~(\ref{F_0}), the gradient force now becomes \begin{equation}
\widetilde{F}_o(\Omega) = f_o(\Omega) \widetilde{x}_b(\Omega) + \widetilde{F}_e(\Omega), \label{F0_Omega_Modulated}
\end{equation}
where $f_o(\Omega)$ is given by Eq.~(\ref{F0_Omega}) and $\widetilde{F}_e(\Omega)$ represents the force component introduced by the input modulation. It is given by the following form:
\begin{equation}
\widetilde{F}_e (\Omega) = \frac{i \sqrt{\Gamma_e} g_{om}}{\omega_0} \left[ \frac{a_0^* \delta \widetilde{A}(\Omega)}{i(\Delta + \Omega) - \Gamma_t/2} + \frac{a_0 \delta \widetilde{A}^*(-\Omega)}{i(\Delta - \Omega) + \Gamma_t/2} \right]. \label{F_e_Omega}
\end{equation}
In particular, in the sideband-unresolved regime, Eq.~(\ref{F_e_Omega}) can be well approximated by
\begin{equation}
\widetilde{F}_e (\Omega) \approx \frac{i \sqrt{\Gamma_e} g_{om}}{\omega_0 (i \Delta - \Gamma_t/2)} \left[ a_0^* \delta \widetilde{A}(\Omega) + a_0 \delta \widetilde{A}^*(-\Omega) \right]. \label{F_e_Omega_Approx}
\end{equation}

In the case that the mechanical excitation is dominated by the external optical modulation, the thermal excitation from the reservoir is negligible and Eqs.~(\ref{dx1_dt}) and (\ref{dx2_dt}) become
\begin{eqnarray}
\frac{d^2 x_b}{dt^2} + \Gamma_{mb} \frac{d x_b}{dt} + \Omega_{mb}^2 x_b + \frac{\kappa}{m_b} x_d &=& \frac{F_o}{m_b}, \label{dx1_dt_Modulated}\\
\frac{d^2 x_d}{dt^2} + \Gamma_{md} \frac{d x_d}{dt} + \Omega_{md}^2 x_d + \frac{\kappa}{m_d} x_b &=& 0. \label{dx2_dt_Modulated}
\end{eqnarray}
Using Eqs.~(\ref{Da_Omega_Modulated}) and (\ref{F0_Omega_Modulated}), following a similar procedure as the previous section, we find that the mechanical displacement for the optically bright mode is now given by
\begin{equation}
\widetilde{x}_b(\Omega) = \frac{\widetilde{F}_e (\Omega)}{m_b} \frac{L_d(\Omega)}{L_b(\Omega) L_d(\Omega) - \eta^4}, \label{x1_Omega_Modulated}
\end{equation}
where $L_b(\Omega)$ and $L_d(\Omega)$ are given in the previous section. Clearly, the mechanical response given in Eq.~(\ref{x1_Omega_Modulated}) is directly analogous to the atomic response in EIT systems \cite{Imamoglu05}.

\section{Correspondence of cavity optomechanics to coherent Stokes and anti-Stokes Raman scattering}
\label{AppG}
In this section, we show a direct correspondence between cavity optomechanics and coherent Stokes and anti-Stokes Raman scattering. The system Hamiltonian of an optomechanical cavity is given by the following general form:
\begin{equation}
{\cal H} = \hbar \omega_0 a^\dag a + \hbar \Omega_m b^\dag b + \hbar g_{om} x_b a^\dag a, \label{Hamilton_NOMS}
\end{equation}
where $a$ and $b$ are the annihilation operators for photon and phonon, respectively, normalized such that $a^\dag a$ and $b^\dag b$ represent the operators for photon and phonon number. $x_b$ is the mechanical displacement for the optically bright mode, related to $b$ by
\begin{equation}
x_b = \sqrt{\frac{\hbar}{2 m_b \Omega_{mb}}} \left(b + b^\dag \right). \label{x_S_operator}
\end{equation}
Therefore, the interaction Hamiltonian between the optical wave and the mechanical motion is given by
\begin{equation}
{\cal H}_i = h g a^\dag a \left( b + b^\dag \right), \label{H_i}
\end{equation}
where the factor $g \equiv \left( \frac{g_{om}^2 \hbar^3}{2 m_b \Omega_{mb}} \right)^{1/2}$.

The mechanical motion modulates the intracavity field to create two optical sidebands. As a result, the optical field can be written as
\begin{equation}
a = a_p + a_s e^{-i \Omega_{mb} t} + a_i e^{i \Omega_{mb} t}, \label{a_decomposed}
\end{equation}
where $a_p$ is the field amplitude of the fundamental wave, and $a_s$ and $a_i$ are those of the generated Stokes and anti-Stokes wave, respectively. As the magnitudes of the Stokes and anti-Stokes sidebands are much smaller than the fundamental wave, when we substitute Eq.~(\ref{a_decomposed}) into Eq.~(\ref{H_i}) and leave only the first-order terms of $a_s$ and $a_i$, under the rotating-wave approximation, the interaction Hamiltonian becomes
\begin{equation}
{\cal H}_i = \hbar g \left( b + b^\dag \right) a_p^\dag a_p + \hbar g b^\dag \left( a_s^\dag a_p + a_p^\dag a_i \right) + \hbar g b \left( a_p^\dag a_s + a_i^\dag a_p \right). \label{H_i_approx}
\end{equation}
In Eq.~(\ref{H_i_approx}), the first term describes the static mechanical actuation, which changes only the equilibrium position of mechanical motion and is neglected in the following analysis, as discussed previously. The second and third terms show clearly that the process corresponds directly to coherent Stokes and anti-Stokes Raman scattering as shown in Fig.~4d in the main text.

\section{The mechanical response with three mode coupling}
\label{AppH}
The coherent mixing of mechanical excitation is universal to gradient-force-based NOMS with a giant optical spring effect. Similar phenomena to that presented for double-disks were also observed in the coupled photonic crystal nanobeams. However, due to its device geometry, the coupled nanobeam has more complex mechanical mode families in which all the even-order mechanical modes are optically dark, because they exhibit a mechanical node at the beam center where the optical mode is located. As the same-order common and differential motions of the two beams have similar mechanical frequencies, they can simultaneously couple to an optically bright mode, leading to multiple excitation interferences on the mechanical response.

In the case when the optically bright mode is coupled to two optically dark modes, the Hamiltonian for the mechanical system is given by the following general form:
\begin{equation}
{\cal H}_m = \sum_{i=b,1,2} {\left( \frac{p_i^2}{2 m_i} + \frac{1}{2} k_i x_i^2 \right)} + \kappa_{1} x_b x_1 + \kappa_{2} x_b x_2, \label{H_m2}
\end{equation}
where $i=b,1,2$ corresponds to the optically bright mode and optically dark modes 1 and 2, respectively. With this Hamiltonian, counting in both the optical gradient force and the Langevin forces from the thermal reservoir, we obtain the equations of motions for the three modes:
\begin{eqnarray}
\frac{d^2 x_b}{dt^2} + \Gamma_{mb} \frac{d x_b}{dt} + \Omega_{mb}^2 x_b + \frac{\kappa_{1}}{m_b} x_1 + \frac{\kappa_{2}}{m_b} x_2 &=& \frac{F_b}{m_b} + \frac{F_o}{m_b}, \label{dx1_dt2}\\
\frac{d^2 x_1}{dt^2} + \Gamma_{m1} \frac{d x_1}{dt} + \Omega_{m1}^2 x_2 + \frac{\kappa_{1}}{m_1} x_b &=& \frac{F_1}{m_1}, \label{dx2_dt2}\\
\frac{d^2 x_2}{dt^2} + \Gamma_{m2} \frac{d x_2}{dt} + \Omega_{m2}^2 x_3 + \frac{\kappa_{2}}{m_2} x_b &=& \frac{F_2}{m_2}, \label{dx3_dt2}
\end{eqnarray}
where the gradient force $F_o$ is given by Eq.~(\ref{F_0}), and the statistical properties of the Langevin forces are given by Eq.~(\ref{F_corr}). Following the same analysis as the previous section, we can obtain the spectral intensity for the mechanical displacement of the optically bright mode as
\begin{equation}
S_{x_b}(\Omega) = \frac{2 k_B T}{m_b} \frac{\eta_{1}^4 \Gamma_{m1} \left|L_2(\Omega) \right|^2 + \eta_{2}^4 \Gamma_{m2} \left|L_1(\Omega) \right|^2 + \Gamma_{mb} \left| L_1(\Omega) L_2(\Omega) \right|^2 }{\left| L_b(\Omega) L_1(\Omega) L_2(\Omega)  - \eta_{1}^4 L_2(\Omega) - \eta_{2}^4 L_1(\Omega) \right|^2}, \label{S_x2}
\end{equation}
where $\eta_{j}^4 \equiv \frac{\kappa_{j}^2}{m_b m_j}$ ($j=1,2$) represents the mechanical coupling coefficient. $L_j(\Omega) = \Omega_{mj}^2 - \Omega^2 - i \Gamma_{mj} \Omega$ ($j=1,2$) and $L_b(\Omega)$ is given by Eq.~(\ref{L1_New}) with $\Omega_{mb}'$ and $\Gamma_{mb}'$ given in Eqs.~(\ref{Omega_mNew}) and (\ref{Gamma_mNew}), respectively. As the optical wave is coupled to the optically bright mode only, the power spectral density of the cavity transmission is still given by Eq.~(\ref{S_P}), with the mechanical response $S_{x_b}$ given in Eq.~(\ref{S_x2}).

\begin{figure}[btp]
\includegraphics[width=0.75\columnwidth]{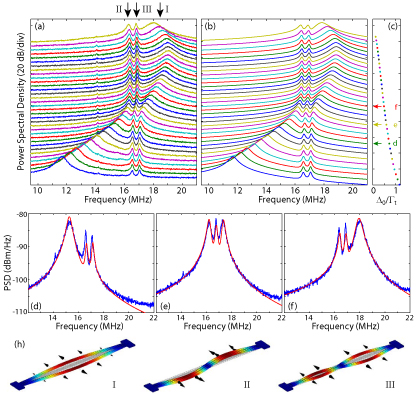}
\caption{\label{Fig5} \scriptsize (a) Experimentally recorded power spectral densities of the cavity transmission for the zipper cavity described in the main text, with an input power of 5.1~mW. Each curve corresponds to a laser frequency detuning indicated in (c). Each curve is relatively shifted by 5 dB in the vertical axis for a better vision of the mechanical frequency tuning and the induced mechanical interference. The optically dark mode II and III have a full-width at half maximum (FWHM) of 0.16 and 0.15~MHz, respectively. The optically bright mode I has an intrinsic FWHM of 0.30~MHz. (b) The corresponding theoretical spectra of the power spectral density. (d)-(f) The detailed spectra of the power spectral density at three frequency detunings indicated by the three arrows in (c). The blue and red curves show the experimental and theoretical spectra, respectively. (h) FEM simulated mechanical motions for the fundamental \emph{differential} mode (I), the second-order \emph{common} (II) and \emph{differential} (III) modes, whose frequencies are indicated by the arrows in (a). The color map indicates the relative magnitude (exaggerated) of the mechanical displacement. }
\end{figure}

Figure \ref{Fig5} shows the PSD of the cavity transmission by launching a continuous wave into a resonance of the coupled nanobeams with an intrinsic and loaded Q factor of $3.0 \times 10^4$ and $2.8 \times 10^4$, respectively. Three mechanical modes are clearly visible, where mode I is the fundamental differential mode [Fig.~\ref{Fig5}(h)I], and mode II and III correspond to the second-order common and differential modes [Fig.~\ref{Fig5}(h)II and III], respectively. Similar to the double-disk NOMS, the gigantic optical spring effect shifts the frequency of the optically bright mode I from its intrinsic value of 8.06~MHz to 19~MHz, crossing over both optically dark modes II and III closely located at 16.54 and 17.04~MHz and resulting in complex interferences on the power spectra [Fig.~\ref{Fig5}(a)]. Equation (\ref{S_x2}) provides an accurate description of the observed phenomena, as shown clearly in Fig.~\ref{Fig5}(b), (d)-(f). Fitting of the PSD results in mechanical coupling coefficients of $\eta_{1} = 3.45$~MHz and $\eta_{2} = 3.48$~MHz, implying that the two optically dark modes couple to the fundamental optically bright mode with a similar magnitude.

\section{Renormalization of mechanical modes by the gradient force}
\label{AppI}
In addition to the gradient force and Langevin force, mechanical motion of the two microdisks (or two nanobeams) changes the gap between them and thus introduces a pressure differential between the gap and the outer region~\cite{Bao07}, which functions as a viscous force to damp the mechanical motion. In the previous sections, we have incorporated this squeeze-film effect in the mechanical damping rate. However, as the generated pressure differential is sensitive to the gap variations, we can expect that the squeeze-film effect behaves quite differently when the two disks/beams vibrate independently or cooperatively. Therefore, we describe it explicitly here for the analysis of mechanical mode renormalization.

In general, the motion of individual disks or nanobeams satisfies the following equations:
\begin{eqnarray}
\frac{d^2 x_1}{dt^2} + \Gamma_{m1} \frac{d x_1}{dt} + \Omega_{m1}^2 x_1 &=& \frac{F_1}{m_1} + \frac{F_o}{m_1} + \frac{F_q}{m_1}, \label{dx1_dt3}\\
\frac{d^2 x_2}{dt^2} + \Gamma_{m2} \frac{d x_2}{dt} + \Omega_{m2}^2 x_2 &=& \frac{F_2}{m_2} - \frac{F_o}{m_2} - \frac{F_q}{m_2}, \label{dx2_dt3}
\end{eqnarray}
where $F_q$ is the viscous force from the squeeze film damping, and $m_j$, $x_j$, $\Omega_{mj}$, $\Gamma_{mj}$, $F_j$ ($j=1,2$) are the effective mass, the mechanical displacement, resonance frequency, damping rate, and the Langevin force for individual disks (or beams), respectively.

The optically bright mechanical mode corresponds to the differential motion of the two disks/beams, with a mechanical displacement given by $x_b \equiv x_1 - x_2$. By transferring Eqs.~(\ref{dx1_dt3}) and (\ref{dx2_dt3}) into the frequency domain, it is easy to find that the mechanical displacement of the optically bright mode is given by
\begin{equation}
\widetilde{x}_b(\Omega) = \frac{ \widetilde{F}_1(\Omega)}{m_1 L_1(\Omega)} - \frac{\widetilde{F}_2(\Omega)}{m_2 L_2(\Omega)} + \left[ \frac{1}{m_1 L_1(\Omega)} + \frac{1}{m_2 L_2(\Omega)} \right] \left[ \widetilde{F}_q(\Omega) + \widetilde{F}_o(\Omega) \right], \label{X_b_Diff}
\end{equation}
where $L_j(\Omega) = \Omega_{mj}^2 - \Omega^2 - i \Gamma_{mj} \Omega$ ($j=1,2$). The squeeze-film effect is produced by the pressure differential between the gap and the outer region introduced by the differential mechanical motion, and thus has a magnitude linearly proportional to the differential displacement. In general, it can be described by $\widetilde{F}_q(\Omega) = f_q(\Omega) \widetilde{x}_b(\Omega)$, where $f_q(\Omega)$ represents the spectral response of the squeeze gas film \cite{Bao07}. Using this form together with Eq.~(\ref{F0_Omega}) in Eq.~(\ref{X_b_Diff}), we obtain the spectral intensity of the optically bright mode displacement,
\begin{equation}
S_{x_b}(\Omega) = \frac{2 k_B T \left[ \frac{\Gamma_{m1}}{m_1} \left| L_2(\Omega) \right|^2 + \frac{\Gamma_{m2}}{m_2} \left| L_1(\Omega) \right|^2 \right]}{\left| L_1(\Omega) L_2(\Omega) - \left[ f_o(\Omega) + f_q(\Omega) \right] \left[ \frac{L_1(\Omega)}{m_2} + \frac{L_2(\Omega)}{m_1} \right] \right|^2}. \label{S_x_b_Renorm}
\end{equation}

As the squeeze-film effect primarily damps the differential motion, its spectral response can be approximated as $f_q(\Omega) \approx i \alpha_q \Omega$. Moreover, since the two disks or nanobeams generally have only slight asymmetry due to fabrication imperfections, they generally have quite close effective masses and energy damping rates: $m_1 \approx m_2 = 2 m_b$ and $\Gamma_{m1} \approx \Gamma_{m2} \equiv \Gamma_m$, where we have used the fact that the effective motional mass of the differential motion is given by $m_b = m_1 m_2/(m_1 + m_2)$. As a result, Eq.~(\ref{S_x_b_Renorm}) can be well approximated by
\begin{equation}
S_{x_b}(\Omega) \approx \frac{k_B T \Gamma_m}{m_b} \frac{\left| L_1(\Omega) \right|^2 + \left| L_2(\Omega) \right|^2 }{\left| L_1(\Omega) L_2(\Omega) - \frac{1}{2} \left[ f_o(\Omega)/m_b + i \Gamma_q \Omega \right] \left[ L_1(\Omega) + L_2(\Omega) \right] \right|^2}, \label{S_x_b_Renorm_approx}
\end{equation}
where $\Gamma_q \equiv \alpha_q/m_b$ represents the damping rate introduced by the squeeze gas film, and the spectral response of the gradient force $f_o(\Omega)$ is given by Eq.~(\ref{F0_Omega}).

The intrinsic mechanical frequencies of 7.790 and 7.995~MHz for the two individual nanobeams are measured from the experimental recorded PSD with a large laser-cavity detuning. The optomechanical coupling coefficient is $68~{\rm GHz/nm}$ and the effective mass is 10.75~pg for the fundamental differential mode, both obtained from FEM simulations (note that these values are different than those quoted in Ref.~\cite{Painter09} due to the different definition of mode amplitude for $x_{b}$). The intrinsic and loaded optical Q factors are $3.0 \times 10^4$ and $2.8 \times 10^4$, respectively, obtained from optical characterization of the cavity resonance. By using these values in Eqs.~(\ref{S_x_b_Renorm_approx}) and (\ref{F0_Omega}), we can easily find the mechanical frequencies and linewidths for the two renormalized modes, where we treat the intrinsic mechanical damping rate $\Gamma_m$ and the squeeze-film-induced damping rate $\Gamma_q$ as fitting parameters. As shown in Fig.~2 of the main text, the theory provides an accurate description of the mechanical mode renormalization, with a fitted intrinsic mechanical and squeeze-film damping rate of 0.03 and 0.2~MHz, respectively.

Similarly, we can obtain the spectral intensity of $x_d \equiv x_1 + x_2$ for the optically-dark mechanical mode, which is given by the following form:
\begin{equation}
S_{x_d}(\Omega) = 2 k_B T \frac{ \frac{\Gamma_{m2}}{m_2} \left| L_1(\Omega) - \frac{2}{m_1} \left[ f_o(\Omega) + f_q(\Omega) \right] \right|^2 + \frac{\Gamma_{m1}}{m_1} \left| L_2(\Omega) - \frac{2}{m_2} \left[ f_o(\Omega) + f_q(\Omega) \right] \right|^2 }{\left| L_1(\Omega) L_2(\Omega) - \left[ f_o(\Omega) + f_q(\Omega) \right] \left[ \frac{L_1(\Omega)}{m_2} + \frac{L_2(\Omega)}{m_1} \right] \right|^2}. \label{S_x_d_Renorm}
\end{equation}
Similar to the optically-bright mode, with $m_1 \approx m_2 = 2 m_b$ and $\Gamma_{m1} \approx \Gamma_{m2} \equiv \Gamma_m$, Eq.~(\ref{S_x_d_Renorm}) can be well approximated by 
\begin{equation}
S_{x_d}(\Omega) \approx \frac{k_B T \Gamma_m}{m_d} \frac{\left| L_1(\Omega) - h(\Omega) \right|^2 + \left| L_2(\Omega) - h(\Omega) \right|^2 }{\left| L_1(\Omega) L_2(\Omega) - \frac{1}{2} h(\Omega) \left[ L_1(\Omega) + L_2(\Omega) \right] \right|^2}, \label{S_x_d_Renorm_approx}
\end{equation}
where $m_d = m/2$ is the effective mass of the common mode and $h(\Omega) \equiv \left[ f_o(\Omega)/m_b + i \Gamma_q \Omega \right]$ represents the total spectral response of the optical gradient force and squeeze film damping. In particular, when the optical-spring-induced frequency shift is much larger than the intrinsic mechanical frequency splitting, the spectral intensities of these two modes reduce to
\begin{equation}
S_{x_b}(\Omega) \approx \frac{2 k_B T \Gamma_m /m_b}{\left| L_o(\Omega) - h(\Omega)\right|^2}, \qquad S_{x_d}(\Omega) \approx \frac{2 k_B T \Gamma_m /m_d}{\left| L_o(\Omega)\right|^2} \label{S_x_b_Reduced}
\end{equation}
where $L_0(\Omega) = (\Omega_{m1} + \Omega_{m2})^2/4 - \Omega^2 - i \Gamma_m \Omega$. Equation (\ref{S_x_b_Reduced}) indicates that the optically bright and dark modes reduce to a pure differential and common modes, respectively.


\begin{thebibliography}{10}
\expandafter\ifx\csname url\endcsname\relax
  \def\url#1{\texttt{#1}}\fi
\expandafter\ifx\csname urlprefix\endcsname\relax\def\urlprefix{URL }\fi
\providecommand{\bibinfo}[2]{#2}
\providecommand{\eprint}[2][]{\url{#2}}

\bibitem{Fano61}
\bibinfo{author}{Fano, U.}
\newblock \bibinfo{title}{Effects of configuration interaction on intensities
  and phase shifts}.
\newblock \emph{\bibinfo{journal}{Phys. Rev.}} \textbf{\bibinfo{volume}{124}},
  \bibinfo{pages}{1866} (\bibinfo{year}{1961}).

\bibitem{Harris90}
\bibinfo{author}{Harris, S.~E.}, \bibinfo{author}{Field, J.~E.} \&
  \bibinfo{author}{Imamo\v{g}lu, A.}
\newblock \bibinfo{title}{Nonlinear optical processes using electromagnetically
  induced transparency}.
\newblock \emph{\bibinfo{journal}{Phys. Rev. Lett.}}
  \textbf{\bibinfo{volume}{64}}, \bibinfo{pages}{1107} (\bibinfo{year}{1990}).

\bibitem{Faist97}
\bibinfo{author}{Faist, J.}, \bibinfo{author}{Capasso, F.},
  \bibinfo{author}{Sitori, C.}, \bibinfo{author}{West, K.~W.} \&
  \bibinfo{author}{Pfeiffer, L.~N.}
\newblock \bibinfo{title}{Controlling the sign of quantum interference by
  tunnelling from quantum wells}.
\newblock \emph{\bibinfo{journal}{Nature}} \textbf{\bibinfo{volume}{390}},
  \bibinfo{pages}{589} (\bibinfo{year}{1997}).

\bibitem{Kroner08}
\bibinfo{author}{Kroner, M.} \emph{et~al.}
\newblock \bibinfo{title}{The nonlinear fano effect}.
\newblock \emph{\bibinfo{journal}{Nature}} \textbf{\bibinfo{volume}{451}},
  \bibinfo{pages}{311} (\bibinfo{year}{2008}).

\bibitem{Scott74}
\bibinfo{author}{Scott, J.~F.}
\newblock \bibinfo{title}{Soft-mode spectroscopy: experimental studies of
  structural phase transitions}.
\newblock \emph{\bibinfo{journal}{Rev. Mod. Phys.}}
  \textbf{\bibinfo{volume}{46}}, \bibinfo{pages}{83} (\bibinfo{year}{1974}).

\bibitem{Hase06}
\bibinfo{author}{Hase, M.}, \bibinfo{author}{Demsar, J.} \&
  \bibinfo{author}{Kitajima, M.}
\newblock \bibinfo{title}{Photoinduced fano resonance of coherent phonons in
  zinc}.
\newblock \emph{\bibinfo{journal}{Phys. Rev. B}} \textbf{\bibinfo{volume}{74}},
  \bibinfo{pages}{212301} (\bibinfo{year}{2006}).

\bibitem{Harris89}
\bibinfo{author}{Harris, S.~E.}
\newblock \bibinfo{title}{Lasers without inversion: interference of
  lifetime-broadened resonances}.
\newblock \emph{\bibinfo{journal}{Phys. Rev. Lett.}}
  \textbf{\bibinfo{volume}{62}}, \bibinfo{pages}{1033} (\bibinfo{year}{1989}).

\bibitem{Imamoglu99}
\bibinfo{author}{Nikonov, D.~E.}, \bibinfo{author}{Imamoglu, A.} \&
  \bibinfo{author}{Scully, M.~O.}
\newblock \bibinfo{title}{Lasers without inversion: interference of
  lifetime-broadened resonances}.
\newblock \emph{\bibinfo{journal}{Phys. Rev. B}} \textbf{\bibinfo{volume}{59}},
  \bibinfo{pages}{12212} (\bibinfo{year}{1999}).

\bibitem{Fan02}
\bibinfo{author}{Fan, S.}
\newblock \bibinfo{title}{Sharp asymmetric line shapes in side-coupled
  waveguide-cavity systems}.
\newblock \emph{\bibinfo{journal}{Appl. Phys. Lett.}}
  \textbf{\bibinfo{volume}{80}}, \bibinfo{pages}{908} (\bibinfo{year}{2002}).

\bibitem{Boyd04}
\bibinfo{author}{Smith, D.~D.}, \bibinfo{author}{Chang, H.},
  \bibinfo{author}{Fuller, K.~A.}, \bibinfo{author}{Rosenberger, A.~T.} \&
  \bibinfo{author}{Boyd, R.~W.}
\newblock \bibinfo{title}{Coupled-resonator-induced transparency}.
\newblock \emph{\bibinfo{journal}{Phys. Rev. A}} \textbf{\bibinfo{volume}{69}},
  \bibinfo{pages}{063804} (\bibinfo{year}{2004}).

\bibitem{Xu06}
\bibinfo{author}{Xu, Q.} \emph{et~al.}
\newblock \bibinfo{title}{Experimental realization of an on-chip all-optical
  analogue to electromagnetically induced transparency}.
\newblock \emph{\bibinfo{journal}{Phys. Rev. Lett.}}
  \textbf{\bibinfo{volume}{96}}, \bibinfo{pages}{123901}
  (\bibinfo{year}{2006}).

\bibitem{Totsuka07}
\bibinfo{author}{Totsuka, K.}, \bibinfo{author}{Kobayashi, N.} \&
  \bibinfo{author}{Tomita, M.}
\newblock \bibinfo{title}{Coupled-resonator-induced transparency}.
\newblock \emph{\bibinfo{journal}{Phys. Rev. Lett.}}
  \textbf{\bibinfo{volume}{98}}, \bibinfo{pages}{213904}
  (\bibinfo{year}{2007}).

\bibitem{Giessen09}
\bibinfo{author}{Liu, N.} \emph{et~al.}
\newblock \bibinfo{title}{Plasmonic analogue of electromagnetically induced
  transparency at the drude damping limit}.
\newblock \emph{\bibinfo{journal}{Nature Materials}}
  \bibinfo{pages}{DOI:10.1038} (\bibinfo{year}{2009}).

\bibitem{Kippenberg05}
\bibinfo{author}{Kippenberg, T.~J.}, \bibinfo{author}{Rokhsari, H.},
  \bibinfo{author}{Carmon, T.}, \bibinfo{author}{Scherer, A.} \&
  \bibinfo{author}{Vahala, K.~J.}
\newblock \bibinfo{title}{{Analysis of Radiation-Pressure Induced Mechanical
  Oscillation of an Optical Microcavity}}.
\newblock \emph{\bibinfo{journal}{Phys. Rev. Lett.}}
  \textbf{\bibinfo{volume}{95}}, \bibinfo{pages}{033901}
  (\bibinfo{year}{2005}).

\bibitem{Gigan06}
\bibinfo{author}{Gigan, S.} \emph{et~al.}
\newblock \bibinfo{title}{{Self-cooling of a micromirror by radiation
  pressure}}.
\newblock \emph{\bibinfo{journal}{Nature}} \textbf{\bibinfo{volume}{444}},
  \bibinfo{pages}{67--70} (\bibinfo{year}{2006}).

\bibitem{Arcizet06}
\bibinfo{author}{Arcizet, O.}, \bibinfo{author}{Cohadon, P.-F.},
  \bibinfo{author}{Briant, T.}, \bibinfo{author}{Pinard, M.} \&
  \bibinfo{author}{Heidmann, A.}
\newblock \bibinfo{title}{{Radiation-pressure cooling and optomechanical
  instability of a micromirror}}.
\newblock \emph{\bibinfo{journal}{Nature}} \textbf{\bibinfo{volume}{444}},
  \bibinfo{pages}{71--73} (\bibinfo{year}{2006}).

\bibitem{Kleckner06}
\bibinfo{author}{Kleckner, D.} \& \bibinfo{author}{Bouwmeester, D.}
\newblock \bibinfo{title}{{Sub-kelvin optical cooling of a micromechanical
  resonator}}.
\newblock \emph{\bibinfo{journal}{Nature}} \textbf{\bibinfo{volume}{444}},
  \bibinfo{pages}{75--78} (\bibinfo{year}{2006}).

\bibitem{Schliesser06}
\bibinfo{author}{Schliesser, A.}, \bibinfo{author}{Del'Haye, P.},
  \bibinfo{author}{Nooshi, N.}, \bibinfo{author}{Vahala, K.~J.} \&
  \bibinfo{author}{Kippenberg, T.~J.}
\newblock \bibinfo{title}{{Radiation Pressure Cooling of a Micromechanical
  Oscillator Using Dynamical Backaction}}.
\newblock \emph{\bibinfo{journal}{Phys. Rev. Lett.}}
  \textbf{\bibinfo{volume}{97}}, \bibinfo{pages}{243905}
  (\bibinfo{year}{2006}).

\bibitem{Povinelli051}
\bibinfo{author}{Povinelli, M.~L.} \emph{et~al.}
\newblock \bibinfo{title}{{Evanescent-wave bonding between optical
  waveguides}}.
\newblock \emph{\bibinfo{journal}{Opt. Lett.}} \textbf{\bibinfo{volume}{30}},
  \bibinfo{pages}{3042--3044} (\bibinfo{year}{2005}).

\bibitem{Eichenfield07}
\bibinfo{author}{Eichenfield, M.}, \bibinfo{author}{Michael, C.~P.},
  \bibinfo{author}{Perahia, R.} \& \bibinfo{author}{Painter, O.}
\newblock \bibinfo{title}{Actuation of micro-optomechanical systems via
  cavity-enhanced optical dipole forces}.
\newblock \emph{\bibinfo{journal}{Nature Photonics}}
  \textbf{\bibinfo{volume}{1}}, \bibinfo{pages}{416} (\bibinfo{year}{2007}).

\bibitem{Li08}
\bibinfo{author}{Li, M.} \emph{et~al.}
\newblock \bibinfo{title}{{Harnessing optical forces in integrated photonic
  cicruits}}.
\newblock \emph{\bibinfo{journal}{Nature}} \textbf{\bibinfo{volume}{456}},
  \bibinfo{pages}{480--484} (\bibinfo{year}{2008}).

\bibitem{Painter09}
\bibinfo{author}{Eichenfield, M.}, \bibinfo{author}{Camacho, R.},
  \bibinfo{author}{Chan, J.}, \bibinfo{author}{Vahala, K.~J.} \&
  \bibinfo{author}{Painter, O.}
\newblock \bibinfo{title}{A picogram- and nanometre-scale photonic-crystal
  optomechanical cavity}.
\newblock \emph{\bibinfo{journal}{Nature}} \textbf{\bibinfo{volume}{459}},
  \bibinfo{pages}{550--555} (\bibinfo{year}{2009}).

\bibitem{Lin09}
\bibinfo{author}{Lin, Q.}, \bibinfo{author}{Rosenberg, J.},
  \bibinfo{author}{Jiang, X.}, \bibinfo{author}{Vahala, K.~J.} \&
  \bibinfo{author}{Painter, O.}
\newblock \bibinfo{title}{Mechanical oscillation and cooling actuated by
  optical gradient forces}.
\newblock \emph{\bibinfo{journal}{arXiv: 0905.2716v1}}  (\bibinfo{year}{2009}).

\bibitem{Rosenberg09}
\bibinfo{author}{Rosenberg, J.}, \bibinfo{author}{Lin, Q.},
  \bibinfo{author}{Vahala, K.~J.} \& \bibinfo{author}{Painter, O.}
\newblock \bibinfo{title}{Static and dynamic wavelength routing via the
  gradient optical force}.
\newblock \emph{\bibinfo{journal}{Nat. Photonics}}
  \bibinfo{pages}{doi:10.1038/nphoton.2009.137} (\bibinfo{year}{2009}).

\bibitem{ref:Ashkin1}
\bibinfo{author}{Ashkin, A.}
\newblock \bibinfo{title}{{Histroy of Optical Trapping and Manipulation of
  Small-Neutral Particle, Atoms, and Molecules}}.
\newblock \emph{\bibinfo{journal}{IEEE J. Quan. Elec.}}
  \textbf{\bibinfo{volume}{6}}, \bibinfo{pages}{841--856}
  (\bibinfo{year}{2000}).

\bibitem{Braginsky77}
\bibinfo{author}{Braginski{\u \i}, V.~B.} \& \bibinfo{author}{Manukin, A.~B.}
\newblock \emph{\bibinfo{title}{Measurement of weak forces in physics
  experiments}} (\bibinfo{publisher}{University of Chicago Press},
  \bibinfo{address}{Chicago}, \bibinfo{year}{1977}).

\bibitem{Braginsky92}
\bibinfo{author}{Braginski{\u \i}, V.~B.}, \bibinfo{author}{Khalili, F.~Y.} \&
  \bibinfo{author}{Thorne, K.~S.}
\newblock \emph{\bibinfo{title}{Quantum measurement}}
  (\bibinfo{publisher}{Cambridge University Press},
  \bibinfo{address}{Cambridge}, \bibinfo{year}{1992}).

\bibitem{Sheard04}
\bibinfo{author}{Sheard, B.~S.}, \bibinfo{author}{Gray, M.~B.},
  \bibinfo{author}{Mow-Lowry, C.~M.}, \bibinfo{author}{McClelland, D.~E.} \&
  \bibinfo{author}{Whitcomb, S.~E.}
\newblock \bibinfo{title}{{Observation and characterization of an optical
  spring}}.
\newblock \emph{\bibinfo{journal}{Phys. Rev. A}} \textbf{\bibinfo{volume}{69}},
  \bibinfo{pages}{051801(R)} (\bibinfo{year}{2004}).

\bibitem{Zadeh07}
\bibinfo{author}{Hossein-Zadeh, M.} \& \bibinfo{author}{Vahala, K.~J.}
\newblock \bibinfo{title}{{Observation of optical spring effect in a
  microtoroidal optomechanical resonator}}.
\newblock \emph{\bibinfo{journal}{Opt. Lett.}} \textbf{\bibinfo{volume}{32}},
  \bibinfo{pages}{1611--1613} (\bibinfo{year}{2007}).

\bibitem{Corbitt07}
\bibinfo{author}{Corbitt, T.} \emph{et~al.}
\newblock \bibinfo{title}{{Optical Dilution and Feedback Cooling of a
  Gram-Scale Oscillator to 6.9 mK}}.
\newblock \emph{\bibinfo{journal}{Phys. Rev. Lett.}}
  \textbf{\bibinfo{volume}{99}}, \bibinfo{pages}{160801}
  (\bibinfo{year}{2007}).

\bibitem{ChanJ09}
\bibinfo{author}{Chan, J.}, \bibinfo{author}{Eichenfield, M.},
  \bibinfo{author}{Camacho, R.} \& \bibinfo{author}{Painter, O.}
\newblock \bibinfo{title}{Optical and mechanical design of a ``zipper''
  photonic crystal optomechanical cavity}.
\newblock \emph{\bibinfo{journal}{Opt. Express}} \textbf{\bibinfo{volume}{17}},
  \bibinfo{pages}{3802--3817} (\bibinfo{year}{2009}).

\bibitem{Bao07}
\bibinfo{author}{Bao, M.} \& \bibinfo{author}{Yang, H.}
\newblock \bibinfo{title}{Squeeze film air damping in mems}.
\newblock \emph{\bibinfo{journal}{Sens. Actuators A}}
  \textbf{\bibinfo{volume}{136}}, \bibinfo{pages}{3} (\bibinfo{year}{2007}).

\bibitem{Barker64}
\bibinfo{author}{A.~S.~Barker, J.} \& \bibinfo{author}{Hopfield, J.~J.}
\newblock \bibinfo{title}{Coupled-optical-ponon-mode theory of the infrared
  dispersion in ${\rm batio_3}$ and ${\rm ktao_3}$}.
\newblock \emph{\bibinfo{journal}{Phys. Rev.}} \textbf{\bibinfo{volume}{135}},
  \bibinfo{pages}{A1732} (\bibinfo{year}{1964}).

\bibitem{Porto68}
\bibinfo{author}{Rousseau, D.~L.} \& \bibinfo{author}{Porto, S. P.~S.}
\newblock \bibinfo{title}{Auger-like resonant interference in raman scattering
  from one- and two-phonon states of ${\rm batio_3}$}.
\newblock \emph{\bibinfo{journal}{Phys. Rev. Lett.}}
  \textbf{\bibinfo{volume}{20}}, \bibinfo{pages}{1354} (\bibinfo{year}{1968}).

\bibitem{Scott70}
\bibinfo{author}{Scott, J.~F.}
\newblock \bibinfo{title}{Hybrid phonons and anharmonic interactions in ${\rm
  alpo_4}$}.
\newblock \emph{\bibinfo{journal}{Phys. Rev. Lett.}}
  \textbf{\bibinfo{volume}{24}}, \bibinfo{pages}{1107} (\bibinfo{year}{1970}).

\bibitem{Zawadowski70}
\bibinfo{author}{Zawadowski, A.} \& \bibinfo{author}{Ruvalds, J.}
\newblock \bibinfo{title}{Indirect coupling and antiresonance of two optic
  phonons}.
\newblock \emph{\bibinfo{journal}{Phys. Rev. Lett.}}
  \textbf{\bibinfo{volume}{24}}, \bibinfo{pages}{1111} (\bibinfo{year}{1970}).

\bibitem{Porto74}
\bibinfo{author}{Chaves, A.}, \bibinfo{author}{Katiyar, R.~S.} \&
  \bibinfo{author}{Porto, S. P.~S.}
\newblock \bibinfo{title}{Coupled modes with $a_1$ symmetry in tetragonal ${\rm
  batio_3}$}.
\newblock \emph{\bibinfo{journal}{Phys. Rev. B}} \textbf{\bibinfo{volume}{10}},
  \bibinfo{pages}{3522} (\bibinfo{year}{1974}).

\bibitem{Imamoglu05}
\bibinfo{author}{Fleischhauer, M.}, \bibinfo{author}{Imamoglu, A.} \&
  \bibinfo{author}{Marangos, J.~P.}
\newblock \bibinfo{title}{Electromagnetically induced transparency: Optics in
  coherent media}.
\newblock \emph{\bibinfo{journal}{Rev. Mod. Phys.}}
  \textbf{\bibinfo{volume}{77}}, \bibinfo{pages}{633} (\bibinfo{year}{2005}).

\bibitem{Alzar02}
\bibinfo{author}{Alzar, C. L.~G.}, \bibinfo{author}{Martinez, M. A.~G.} \&
  \bibinfo{author}{Nussenzveig, P.}
\newblock \bibinfo{title}{Classical analog of electromagnetically induced
  transparency}.
\newblock \emph{\bibinfo{journal}{Am. J. Phys.}} \textbf{\bibinfo{volume}{70}},
  \bibinfo{pages}{37} (\bibinfo{year}{2002}).

\bibitem{Hemmer88}
\bibinfo{author}{Hemmer, P.~R.} \& \bibinfo{author}{Prentiss, M.~G.}
\newblock \bibinfo{title}{Coupled-pendulum model of the stimulated resonance
  raman effect}.
\newblock \emph{\bibinfo{journal}{J. Opt. Soc. Am. B}}
  \textbf{\bibinfo{volume}{5}}, \bibinfo{pages}{1613} (\bibinfo{year}{1988}).

\bibitem{Hau01}
\bibinfo{author}{Liu, C.}, \bibinfo{author}{Dutton, Z.},
  \bibinfo{author}{Behroozi, C.~H.} \& \bibinfo{author}{Hau, L.~V.}
\newblock \bibinfo{title}{Observation of coherent optical information storage
  in an atomic medium using halted light pulses}.
\newblock \emph{\bibinfo{journal}{Nature}} \textbf{\bibinfo{volume}{409}},
  \bibinfo{pages}{490} (\bibinfo{year}{2001}).

\bibitem{Lukin03}
\bibinfo{author}{Bajcsy, M.}, \bibinfo{author}{Zibrov, A.~S.} \&
  \bibinfo{author}{Lukin, M.~D.}
\newblock \bibinfo{title}{Stationary pulses of light in an atomic medium}.
\newblock \emph{\bibinfo{journal}{Nature}} \textbf{\bibinfo{volume}{426}},
  \bibinfo{pages}{638} (\bibinfo{year}{2003}).

\bibitem{Fan04}
\bibinfo{author}{Fatih, M.} \& \bibinfo{author}{Fan, S.}
\newblock \bibinfo{title}{Stopping light all optically}.
\newblock \emph{\bibinfo{journal}{Phys. Rev. Lett.}}
  \textbf{\bibinfo{volume}{92}}, \bibinfo{pages}{083901}
  (\bibinfo{year}{2004}).

\bibitem{Fan05}
\bibinfo{author}{Yanik, M.~F.} \& \bibinfo{author}{Fan, S.}
\newblock \bibinfo{title}{Stopping and storing light coherently}.
\newblock \emph{\bibinfo{journal}{Phys. Rev. A}} \textbf{\bibinfo{volume}{71}},
  \bibinfo{pages}{013803} (\bibinfo{year}{2005}).

\bibitem{Anetsberger08}
\bibinfo{author}{Anetsberger, G.}, \bibinfo{author}{Riviere, R.},
  \bibinfo{author}{Schliesser, A.}, \bibinfo{author}{Arcizet, O.} \&
  \bibinfo{author}{Kippenberg, T.~J.}
\newblock \bibinfo{title}{Ultralow-disspation optomechanical resonators on a
  chip}.
\newblock \emph{\bibinfo{journal}{Nat. Photonics}}
  \textbf{\bibinfo{volume}{2}}, \bibinfo{pages}{627--633}
  (\bibinfo{year}{2008}).

\bibitem{Xu07}
\bibinfo{author}{Xu, Q.}, \bibinfo{author}{Dong, P.} \&
  \bibinfo{author}{Lipson, M.}
\newblock \bibinfo{title}{Breaking the delay-bandwidth limit in a photonic
  structure}.
\newblock \emph{\bibinfo{journal}{Nature Phys.}} \textbf{\bibinfo{volume}{3}},
  \bibinfo{pages}{406} (\bibinfo{year}{2007}).

\bibitem{Verbridge081}
\bibinfo{author}{Verbridge, S.~S.}, \bibinfo{author}{Craighead, H.~G.} \&
  \bibinfo{author}{Parpia, J.~M.}
\newblock \bibinfo{title}{A megahertz nanomechanical resonator with room
  temperature quality factor over a million}.
\newblock \emph{\bibinfo{journal}{Appl. Phys. Lett.}}
  \textbf{\bibinfo{volume}{92}}, \bibinfo{pages}{013112}
  (\bibinfo{year}{2008}).

\bibitem{ThompsonJD08}
\bibinfo{author}{Thompson, J.~D.} \emph{et~al.}
\newblock \bibinfo{title}{{Strong dispersive coupling of a high-finesse cavity
  to micromechanical membrane}}.
\newblock \emph{\bibinfo{journal}{Nature}} \textbf{\bibinfo{volume}{452}},
  \bibinfo{pages}{72--75} (\bibinfo{year}{2008}).

\bibitem{Kippenberg03}
\bibinfo{author}{Kippenberg, T.~J.}, \bibinfo{author}{Kalkman, J.},
  \bibinfo{author}{Polman, A.} \& \bibinfo{author}{Vahala, K.~J.}
\newblock \bibinfo{title}{Demonstration of an erbium-doped microdisk laser on a
  silicon chip}.
\newblock \emph{\bibinfo{journal}{Phys. Rev. A}} \textbf{\bibinfo{volume}{74}},
  \bibinfo{pages}{051802(R)} (\bibinfo{year}{2006}).

\end{thebibliography}
\end{document}